\documentclass[aps, prl, twocolumn, showpacs, superscriptaddress, floatfix]{revtex4}

\usepackage{graphicx}
\usepackage{dcolumn}
\usepackage{bm}

\begin{document}

\title{Frustration by competing interactions in the highly-distorted double perovskites La$_2$NaB$'$O$_6$ (B$'$~$=$~Ru, Os)}

\author{A.A. Aczel}
\altaffiliation{author to whom correspondences should be addressed: E-mail:[aczelaa@ornl.gov]}
\affiliation{Quantum Condensed Matter Division, Neutron Sciences Directorate, Oak Ridge National Laboratory, Oak Ridge, TN 37831, USA}
\author{D.E. Bugaris}
\affiliation{Department of Chemistry and Biochemistry, University of South Carolina, Columbia, SC 29208, USA}
\author{L. Li}
\affiliation{Department of Materials Science and Engineering, University of Tennessee, Knoxville, TN 37996, USA}
\author{J.-Q. Yan}
\affiliation{Department of Materials Science and Engineering, University of Tennessee, Knoxville, TN 37996, USA}
\affiliation{Materials Science and Technology Division, Oak Ridge National Laboratory, Oak Ridge, TN 37831, USA}
\author{C. de la Cruz}
\affiliation{Quantum Condensed Matter Division, Neutron Sciences Directorate, Oak Ridge National Laboratory, Oak Ridge, TN 37831, USA}
\author{H.-C. zur Loye}
\affiliation{Department of Chemistry and Biochemistry, University of South Carolina, Columbia, SC 29208, USA}
\author{S.E. Nagler}
\affiliation{Quantum Condensed Matter Division, Neutron Sciences Directorate, Oak Ridge National Laboratory, Oak Ridge, TN 37831, USA}
\affiliation{CIRE, University of Tennessee, Knoxville, TN 37996, USA}

\date{\today}

\begin{abstract}
The usual classical behaviour of $S$~$=$~3/2, B-site ordered double perovskites generally results in simple, commensurate magnetic ground states. In contrast, heat capacity and neutron powder diffraction measurements for the $S$~$=$~3/2 systems La$_2$NaB$'$O$_6$ (B~$=$~Ru, Os) reveal an incommensurate magnetic ground state for La$_2$NaRuO$_6$ and a drastically suppressed ordered moment for La$_2$NaOsO$_6$. This behaviour is attributed to the large monoclinic structural distortions of these double perovskites. The distortions have the effect of weakening the nearest neighbour superexchange interactions, presumably to an energy scale that is comparable to the next nearest neighbour superexchange. The exotic ground states in these materials can then arise from a competition between these two types of antiferromagnetic interactions, providing a novel mechanism for achieving frustration in the double perovskite family. 
\end{abstract}


\maketitle

\renewcommand{\topfraction}{0.85}
\renewcommand{\textfraction}{0.1}
\renewcommand{\floatpagefraction}{0.75}

Insulating, B-site ordered double perovskites of the formula A$_2$BB$'$O$_6$ have attracted considerable interest recently due to the opportunity to study geometric frustration on a face-centered cubic (FCC) lattice. This situation arises when the only magnetic ions in the system can be associated with the B$'$ site and are governed by antiferromagnetic (AF) nearest neighbour (NN) interactions. There should also be minimal site mixing between the B and B$'$ sites. Since the double perovskite structure type is very versatile with respect to chemical substitution and the B sites can accommodate a variety of transition metals, systematic studies can be performed to investigate the effects of changing the spin quantum number $S$ and increasing the relativistic spin-orbit coupling by considering materials with 4$d$ and 5$d$ electrons. A series of exotic magnetic ground states have been observed previously in $S$~$=$~1/2 and 1 systems, including a collective singlet state coexisting with paramagnetism in Ba$_2$YMoO$_6$\cite{10_aharen, 11_carlo}, a collective singlet state in La$_2$LiReO$_6$\cite{10_aharen_2}, spin freezing without long-range order in Ba$_2$YReO$_6$\cite{10_aharen_2}, Sr$_2$MgReO$_6$\cite{03_wiebe} and Sr$_2$CaReO$_6$\cite{02_wiebe}, short-range order in La$_2$LiMoO$_6$\cite{10_aharen}, and a ferromagnetic (FM) Mott insulating state in Ba$_2$NaOsO$_6$\cite{02_stitzer, 07_erickson}. Theoretical studies have also indicated that a wealth of other magnetic ground states are possible in these 4$d$ and 5$d$ quantum spin systems\cite{10_chen, 10_chen_2}.

\begin{figure}
\centering
\scalebox{0.13}{\includegraphics{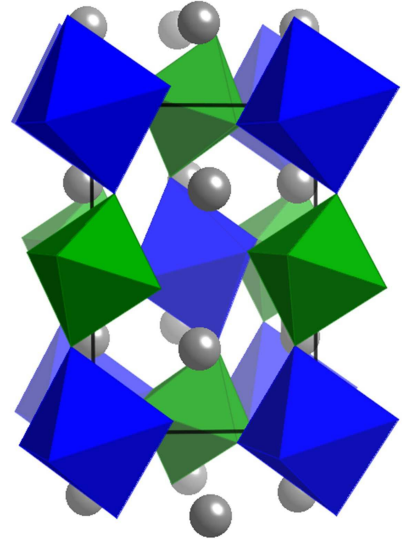}}
\caption{\label{Fig1} Double perovskite structure, with the large blue octahedra representing NaO$_6$, the small green octahedra depicting (Ru,Os)O$_6$, and the isolated grey spheres corresponding to La atoms. The small ionic radius of La leads to a large tilting of the NaO$_6$ and (Ru,Os)O$_6$ octahedra in La$_2$NaRuO$_6$ and La$_2$NaOsO$_6$.}
\end{figure} 

4$d$ and 5$d$ double perovskites with larger $S$~$=$~3/2 and 5/2 spins are expected to behave more classically. NN AF exchange interactions along extended superexchange B$'$-O-O-B$'$ pathways are often dominant in these materials, resulting in Type I AF order, following the notation in Ref. \cite{89_battle}. Since the magnetic atoms are on a geometrically-frustrated FCC lattice, this state is a compromise where eight of the NN spins are AF-aligned, four are FM, and all NNN spins are also FM. There are many examples where this magnetic ground state is realized, including Ca$_2$LaRuO$_6$\cite{83_battle}, Sr$_2$YRuO$_6$\cite{84_battle}, Sr$_2$LuRuO$_6$, Ba$_2$YRuO$_6$, Ba$_2$LuRuO$_6$\cite{89_battle}, La$_2$LiRuO$_6$\cite{03_battle}, and Sr$_2$TeMnO$_6$\cite{06_martin}. On the other hand, if AF next nearest neighbour (NNN) exchange is the most important interaction, these materials do not show strong frustration effects and are often characterized by Type II AF order\cite{89_battle}. The Type II magnetic state ensures that all NNN spins are AF-aligned. Experimental realizations include Ca$_2$WMnO$_6$, Sr$_2$WMnO$_6$, Sr$_2$MoMnO$_6$\cite{02_munoz} and LaANbCoO$_6$ (A~$=$~Ca, Ba, Sr)\cite{04_bos}. Finally, Type III AF order has been observed in the system Ba$_2$LaRuO$_6$\cite{83_battle}. This state arises when the NN AF interaction is large but the NNN AF interaction is not negligible. While the spin alignment between NNs is the same as for Type I AF, two out of the six NNN spins also become AF in the Type III ordered state.     

In the ideal double perovskite cubic structure, both the B and B$'$ sites form FCC lattices. In practice, many double perovskites exhibit structural distortions from the ideal cubic behaviour, most commonly caused by introducing a very small cation into the A site. This atomic position is indicated by the grey isolated spheres in the double perovskite structure as illustrated in Fig.~\ref{Fig1}. The structural distortion often lowers the crystal symmetry from cubic to monoclinic, while the magnitude of the distortion is best described by the deviation of the crystallographic angle $\beta$ from 90$^\circ$. Although the BO$_6$ and the B$'$O$_6$ octahedra of these monoclinic double perovskites remain nearly ideal, the structural distortion tilts them by varying degrees, with the amount of tilting depending on both the size of the A site cation and the difference in the ionic radii of the B and B$'$ sites. Furthermore, the B$'$ magnetic atoms form a pseudo-FCC sublattice characterized by two sets each of two-fold degenerate and four-fold degenerate NN distances, rather than the equivalent 12 NN magnetic atoms inherent to an ideal FCC lattice. The modified structure has important consequences for the magnetic behaviour. In general, the apparent strength of NN interactions is significantly weaker in distorted systems. Moreover, the geometric frustration in materials where the NN interaction dominates is partially relieved. As typical examples, the cubic system Ba$_2$YRuO$_6$ has a Curie-Weiss temperature of -522 K, an ordering temperature of 37 K, and a frustration index of 14, while the same parameters for the monoclinic material La$_2$LiRuO$_6$ were found to be -184 K, 23 K, and 8\cite{10_aharen_3}. 

\begin{figure}
\centering
\scalebox{0.42}{\includegraphics{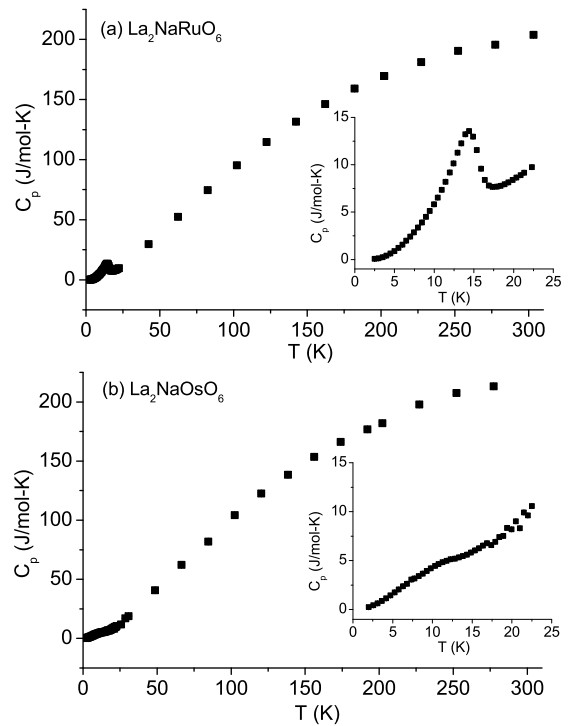}}
\caption{\label{Fig2} Heat capacity measurements for (a) La$_2$NaRuO$_6$ and (b) La$_2$NaOsO$_6$. The insets show the low temperature behaviour for each material. A clear lambda anomaly corresponding to a magnetic transition is visible for La$_2$NaRuO$_6$, while only a broad feature is observed in the La$_2$NaOsO$_6$ data.}
\end{figure} 

The syntheses, crystal structures and magnetic susceptibilities of the monoclinic double perovskites La$_2$NaRuO$_6$ and La$_2$NaOsO$_6$ were recently reported\cite{04_gemmill, 05_gemmill}. These materials combine a small A site cation with a relatively large ionic radii difference between the B and B$'$ sites, and are therefore subjected to very large structural distortions, with $\beta$ angles of 90.495(2)$^\circ$ and 90.587(2)$^\circ$ corresponding to B$'$O$_6$ octahedra tilts of 20.8$^{\circ}$ and 21.0$^{\circ}$ at room temperature. It is interesting to note that the vast majority of monoclinic double perovskites have $\beta$ angles that deviate from 90$^\circ$ by less than 0.3$^\circ$ (see e.g. Ref. \cite{04_bos}). The large structural distortions for  La$_2$NaRuO$_6$ and La$_2$NaOsO$_6$ result in Curie-Weiss temperatures of -67 K and -74 K respectively, consistent with the expected decrease in the strength of the NN exchange interactions compared to cubic double perovskites. The magnetic susceptibility of La$_2$NaRuO$_6$ shows a small downturn at 16~K that has been attributed to antiferromagnetic order. Recent band structure calculations have also been performed and predict that La$_2$NaRuO$_6$ is insulating\cite{08_ahmed}. In contrast, La$_2$NaOsO$_6$ shows the onset of a ferromagnetic-like increase in the magnetic susceptibility at 17~K. A small amount of magnetic hysteresis is also visible in the magnetization as a function of field, although the ferromagnetic component of the moment remains below 0.2~$\mu_B$ in an applied field of 4 T. These properties are certainly not consistent with Type I AF order, and consideration of this combined data set leads one to expect that the ground state is likely a canted antiferromagnet. La$_2$NaOsO$_6$ is also expected to be an insulator based on the large spatial separation of the Os ions, and this behaviour has been confirmed in the related system Ba$_2$NaOsO$_6$\cite{07_erickson}.

In this work, we performed heat capacity and neutron powder diffraction (NPD) measurements to investigate the magnetic ground states of these materials in more detail. Our results reveal a clear lambda anomaly in the specific heat corresponding to incommensurate spin ordering for La$_2$NaRuO$_6$ with an ordering wavevector of (0~0~1$\pm \delta$) with $\delta$~$=$~0.091. The magnetic structure is best explained by two independent, interpenetrating helices made up of alternating ab-planes of spins. Surprisingly, La$_2$NaOsO$_6$ exhibits only a broad, weak feature in the specific heat around 10~K and no magnetic Bragg peaks are detected down to 4 K. This puts an upper limit of 0.2~$\mu_B$ on ordered moments associated with the proposed canted antiferromagnetic ground state. 

\begin{figure*}
\centering
\scalebox{0.65}{\includegraphics{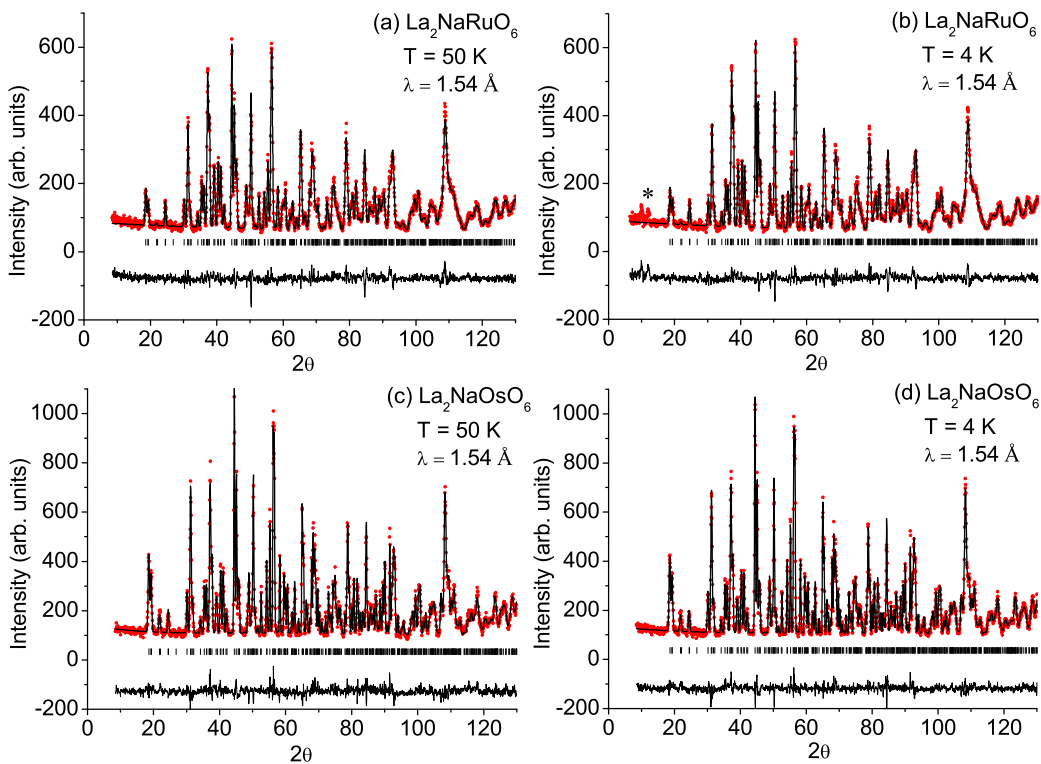}}
\caption{\label{Fig3} Neutron diffraction data with $\lambda$~$=$~1.538~\AA~is shown in (a) and (b) for La$_2$NaRuO$_6$ and (c) and (d) for La$_2$NaOsO$_6$. The asterisk in (b) labels the largest magnetic Bragg peaks observed in the Ru system at low temperature.}
\end{figure*} 
 
Polycrystalline La$_2$NaRuO$_6$ was synthesized via a solid state reaction. La$_2$O$_3$ (Alfa Aesar, 99.99\%) was first activated by heating in air at 1000$^\circ$C for 12 h, Na$_2$CO$_3$ (Mallinckrodt, 99.95\%) was dried overnight at 150$^\circ$C, and RuO$_2$ was prepared by heating Ru (Engelhard, 99.95\%) in air at 1000$^\circ$C for 24 h. The starting materials were then mixed together in a 1:0.55:1 ratio, and this included a 10\% molar excess of Na$_2$CO$_3$ to offset the volatilization of Na$_2$O during heating. This mixture was heated to 500$^\circ$C in 1 h, held at 500$^\circ$C for 8 h, heated to 900$^\circ$C in 1 h, and held at 900$^\circ$C for 12 h before turning off the furnace and allowing the sample to cool to room temperature. The sample identity was confirmed via powder X-ray diffraction, where the data were collected on a Rigaku Ultima IV powder diffractometer using Cu K$\alpha$ radiation. Data were collected on the high-speed D/teX Ultra detector in 0.02$^\circ$ steps over the 2$\theta$ range 10-80$^\circ$ with a speed of 10$^\circ$/min.   

Polycrystalline La$_2$NaOsO$_6$ was also synthesized via a solid state reaction. The starting materials were nearly identical, with the only difference being that Os (J\&J Materials Inc.) replaced RuO$_2$. The heat treatment was also slightly modified, as the starting mixture was heated to 900$^\circ$C in 1.5 h and held at 900$^\circ$ for 12 h before turning off the furnace and allowing the sample to cool to room temperature. Subsequent powder X-ray diffraction revealed an impurity phase of La$_2$O$_3$, so the sample was ground together with additional Na$_2$CO$_3$ and Os and then subjected to the same heating profile as before. This step was repeated one additional time, and then powder X-ray diffraction revealed a single phase sample of La$_2$NaOsO$_6$. 

The heat capacity data was collected in a Physical Property Measurement System using cold-pressed pellets. For the NPD experiment, roughly 5 g of each polycrystalline sample was loaded in a closed-cycle refrigerator and studied using the HB-2A powder diffractometer at the High Flux Isotope Reactor of Oak Ridge National Laboratory. Data from HB-2A were collected with neutron wavelengths $\lambda$~$=$~1.538~\AA~and $\lambda$~$=$~2.41~\AA~at temperatures of 4 - 50 K using a collimation of 12$'$-open-6$'$. The shorter wavelength gives a greater intensity and higher $Q$ coverage that was used to investigate the crystal structures in this low temperature regime, while the longer wavelength gives lower $Q$ coverage and greater resolution that was important for investigating the magnetic structures of these materials. The NPD data was analyzed using the Rietveld refinement program FullProf\cite{93_rodriguez}. 

The heat capacity measurements for both specimens are presented in Fig.~\ref{Fig2} as a function of temperature. The low temperature behaviour is displayed in the insets, and the difference between the two materials is quite striking. The La$_2$NaRuO$_6$ data shows a well-defined lambda anomaly around 15~K; this corresponds well to the magnetic transition temperature inferred from the magnetic susceptibility. However, the specific heat of La$_2$NaOsO$_6$ is only characterized by a weak, broad feature around 10~K and no discernable lambda anomaly is observed.  

\begin{center}
\begin{table}[htb]
\caption{Structural parameters for La$_2$NaRuO$_6$ and La$_2$NaOsO$_6$ at T~$=$~4~K extracted from the $\lambda$~$=$~1.538~\AA~neutron powder diffraction data. } 

\medskip

(a) La$_2$NaRuO$_6$ \\
Space group P2$_1$/n \\ 
a~$=$~5.5956(2)~\AA \\
b~$=$~5.9098(1)~\AA \\ 
c~$=$~8.0013(2)~\AA \\ 
$\beta$~$=$~90.384(2)$^\circ$ \\
$\chi^2$~$=$~2.98 \\
R$_{wp}$~$=$~6.24~\% \\

\begin{tabular}{| c | c | c | c | c |}
\hline 
Atom & Site & x & y & z  \\  \hline
La & 4e & 0.4838(4) & 0.0643(2) & 0.2524(4)    \\  
Na & 2a & 0 & 0 & 0  \\  
Ru & 2b & 0.5 & 0.5 & 0  \\  
O$_1$ & 4e & 0.2108(5) & 0.3228(5) & 0.0475(4) \\  
O$_2$ & 4e & 0.6016(5) & 0.4636(5) & 0.2318(4) \\  
O$_3$ & 4e & 0.3281(6) & 0.7793(5) & 0.0572(4) \\ \hline  
\end{tabular}

\medskip

(b) La$_2$NaOsO$_6$ \\
Space group P2$_1$/n \\ 
a~$=$~5.6062(1)~\AA \\
b~$=$~5.9172(1)~\AA \\ 
c~$=$~8.0296(1)~\AA \\ 
$\beta$~$=$~90.457(2)$^\circ$ \\
$\chi^2$~$=$~3.01 \\
R$_{wp}$~$=$~5.73~\% \\

\begin{tabular}{| c | c | c | c | c |}
\hline 
Atom & Site & x & y & z  \\  \hline
La & 4e & 0.4843(3) & 0.0624(2) & 0.2523(3)    \\  
Na & 2a & 0 & 0 & 0  \\  
Os & 2b & 0.5 & 0.5 & 0  \\  
O$_1$ & 4e & 0.2131(4) & 0.3244(4) & 0.0471(3) \\  
O$_2$ & 4e & 0.6024(4) & 0.4609(4) & 0.2298(3) \\  
O$_3$ & 4e & 0.3314(4) & 0.7780(4) & 0.0588(3) \\  \hline
\end{tabular}
\end{table}
\end{center}

Figure~\ref{Fig3} and Table I show $\lambda$~$=$~1.538~\AA~NPD data for monoclinic La$_2$NaRuO$_6$ and La$_2$NaOsO$_6$ at T~$=$~4~K. Comparing these results with previous X-ray diffraction data collected on single crystals at room temperature\cite{04_gemmill, 05_gemmill} reveals no evidence for a structural phase transition between 300 and 4~K in either material, and the structural distortion remains large at all temperatures as indicated by the 4~K $\beta$ values of 90.384(2)$^\circ$ and 90.457(2)$^\circ$ for the Ru and Os systems respectively. These angles are somewhat smaller than the values determined from the room temperature X-ray measurements, suggesting that these materials become more cubic with decreasing temperature. However, the NPD measurements reveal no change in $\beta$ between 4 K and 50 K within one standard deviation for both systems. The Rietveld refinements also confirm that there is essentially no site mixing between the Na and Ru/Os atomic positions, as expected for double perovskite systems with a charge difference of +4 between the B and B$'$ sites\cite{93_andersen}.

\begin{figure}
\centering
\scalebox{0.36}{\includegraphics{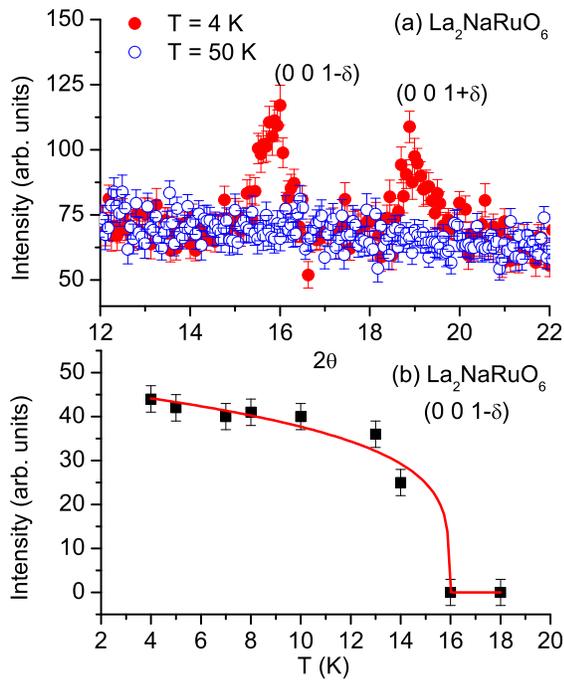}}
\caption{\label{Fig4} (a) $\lambda$~$=$~2.41~\AA~NPD data showing the (0 0 1$\pm \delta$) magnetic peaks that appear at low temperature. (b) Intensity of the incommensurate (0 0 1-$\delta$) peak plotted as a function of temperature, indicating a magnetic transition at T~$=$~16~K in good agreement with magnetic susceptibility\cite{04_gemmill} and heat capacity data. The solid line is a guide to the eye.}
\end{figure}

As indicated by the asterisk in Fig.~\ref{Fig3}(b) and the low angle diffraction plot of the $\lambda$~$=$~2.41~\AA~data shown in Fig.~\ref{Fig4}(a), below $\sim$~16~K additional scattering is observed in the La$_2$NaRuO$_6$ neutron diffraction pattern at the incommensurate positions (0  0 1$\pm\delta$). This is indicative of magnetic order with a propagation vector $\vec{k}$~$=$~$(0~0~0.091)$. Fig.~\ref{Fig4}(b) shows the temperature-dependence of the (0 0 1-$\delta$) magnetic reflection. The incommensurability was found to be roughly temperature-independent and the transition temperature associated with this reflection corresponds well to the value of T~$=$~15-16~K inferred from susceptibility\cite{04_gemmill} and heat capacity measurements. To the best of our knowledge, this is the first time that an incommensurate magnetic ground state has been observed in a B-site ordered double perovskite system with only one type of magnetic atom in the unit cell. In fact, incommensurate magnetic ground states are rarely found in double perovskite systems in general. One of the few known examples is the helical spin structure observed in Ba$_2$CoReO$_6$\cite{73_khattak}, a material with two types of magnetic atoms. 

\begin{figure}
\centering
\scalebox{0.17}{\includegraphics{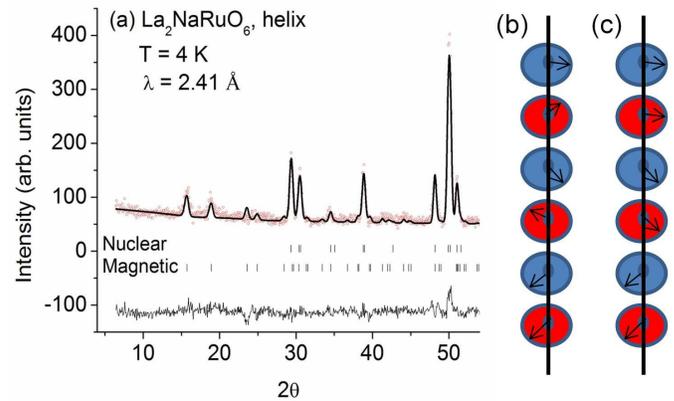}}
\caption{\label{Fig5} (a) La$_2$NaRuO$_6$ Rietveld refinement results with $\lambda$~$=$~2.41~\AA~for a magnetic model of two independent, interpenetrating helices. (b), (c) Two possible helical magnetic structures discussed in the text, with the different colors corresponding to the ab-planes that make up the two helices. Although (c) is shown with no relative phase angle between the helices, this is only for simplicity and is not strictly true. The refinement actually yielded a relative phase angle of 104(5)$^\circ$. }
\end{figure}

The (001) magnetic satellites of La$_2$NaRuO$_6$ are very intense relative to other magnetic peaks including the (100) and (010) satellites, suggesting that the ordered spins align in the ab-plane. The simplest incommensurate magnetic structures consistent with this feature and the observed propagation vector are single helical and sinusoidal spin density wave arrangements. However, these spin configurations yield very little intensity for the satellite peaks around the (001) Bragg position and are instead characterized by strong (002) satellite peaks; both of these features are clearly inconsistent with the data. For this reason, incommensurate magnetic structures were considered that consist of two independent, interpenetrating helices, with every other ab-plane of spins forming a single helix characterized by a turn angle of $\sim$~32.7$^\circ$. The refinement result using the $\lambda$~$=$~2.41~\AA~ data and a model assuming that the helices have the same chirality is shown in Fig.~\ref{Fig5}(a); this magnetic structure reproduces the intensity for the (001) satellite reflections very well. One can also obtain good agreement with the data by using a similar model where the two helices have opposite chirality instead. These two magnetic structures are depicted in Fig.~\ref{Fig5}(b) and (c), with the alternating colors representing the planes forming the two different helices. In both cases, the ordered moment size for Ru was found to be 1.87(7)~$\mu_B$ and this value is consistent with the ordered moments reported for other Ru double perovskites. Since the Ru$^{5+}$ magnetic form factor is not available in the literature, the refinements were attempted with the $<j_0>$ form factors for both Ru$^{+}$ and Os$^{5+}$\cite{11_kobayashi} and the results yielded the same ordered moment sizes for Ru within one standard deviation. Note that magnetic models consisting of two interpenetrating spin density waves can also be used to describe the data, although these spin configurations are rare in insulating magnets.

\begin{figure}
\centering
\scalebox{0.11}{\includegraphics{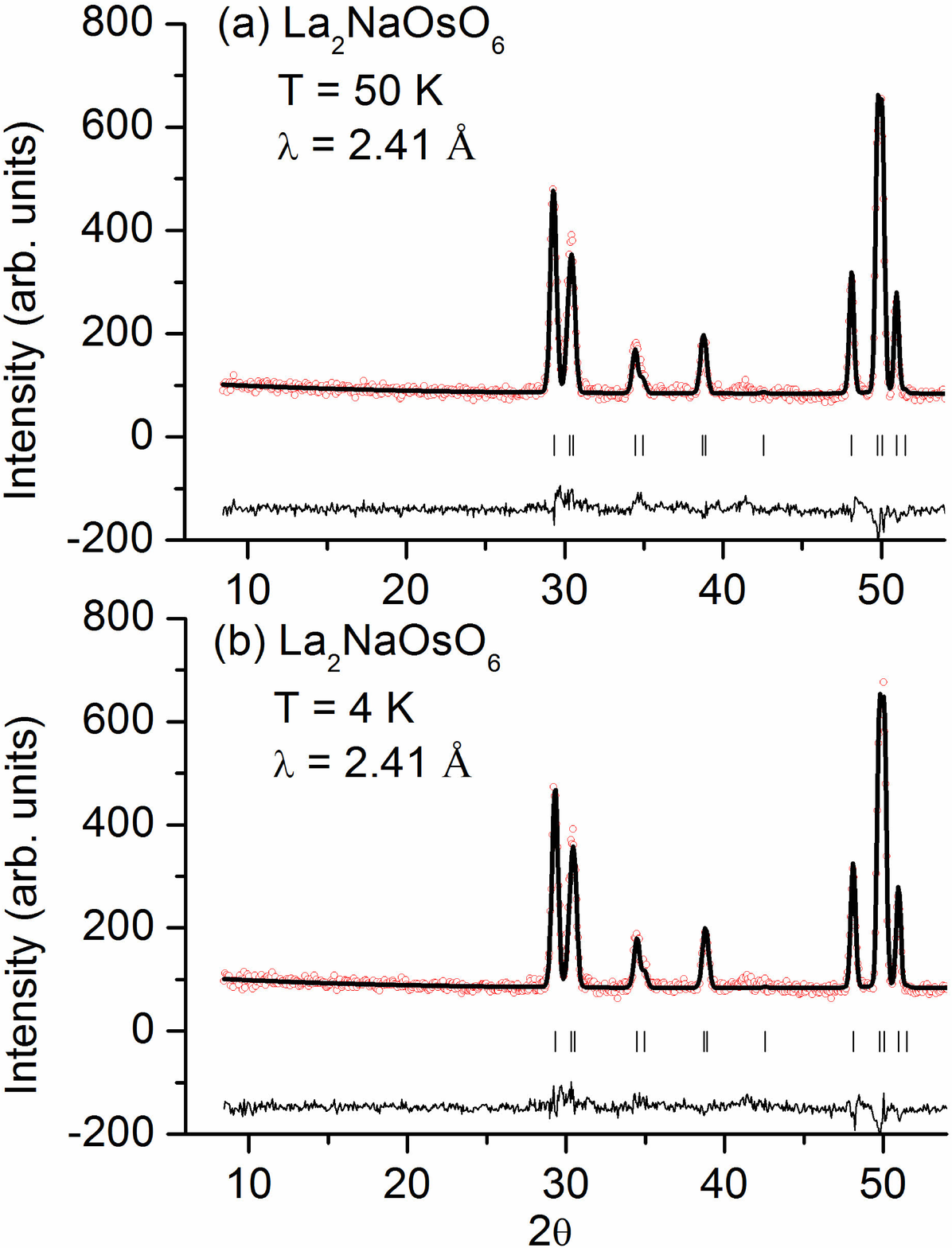}}
\caption{\label{Fig6} La$_2$NaOsO$_6$ neutron diffraction data with $\lambda$~$=$~2.41~\AA, indicating the absence of magnetic peaks down to 4~K.}
\end{figure}

As shown in Fig.~\ref{Fig6}, no additional scattering is observed in the La$_2$NaOsO$_6$ neutron diffraction pattern down to 4~K despite the relatively large Os$^{5+}$ spin $S$~$=$~3/2 and a possible signature of magnetic ordering in the susceptibility data. As outlined in Ref.~\cite{07_erickson} for the related system Ba$_2$NaOsO$_6$, the OsO$_6$ octahedra likely form molecular orbitals due to the comparable energy scales of the Os 5$d$ and O 2$p$ orbitals. The increased covalency resulting from this process is expected to decrease the Os ordered moment. Spin-orbit coupling generally plays an important role in the magnetism of 5$d$ systems also\cite{09_kim}, and is another mechanism that can lead to a reduction in the spin-only value of the ordered moment. Finally, the magnetic Bragg peaks of 4$d$ and 5$d$ systems tend to be weaker than for their 3$d$ counterparts due to the 4$d$ and 5$d$ magnetic form factors decreasing at a faster rate with increasing $Q$. 

To appreciate the magnitude of these different effects on an Os$^{5+}$ material, it is instructive to consider the systems NaOsO$_3$ and Ca$_3$LiOsO$_6$. A combined neutron and X-ray study for NaOsO$_3$\cite{12_calder} revealed an Os ordered moment of 1.0(1)~$\mu_B$, while the magnetic structure of Ca$_3$LiOsO$_6$ was also determined recently\cite{12_calder_2} and found to have an ordered Os moment of 2~-~2.3 $\mu_B$. The local environment of the magnetic atoms in these two materials is very similar to La$_2$NaOsO$_6$. In all three cases, the Os atoms are in slightly-distorted octahedral oxygen cages with nearly identical Os-O bond lengths and no O-Os-O angle deviating from 180$^\circ$ or 90$^\circ$ by more than 1$^\circ$, so any covalency effects that reduce the ordered Os moment in these materials should not be drastically different. Due to the orbital singlet ground states of the Os atoms in these systems, spin-orbit coupling should also have a negligible effect on reducing the ordered moment size. Moreover, the expected canted antiferromagnetic ground state for La$_2$NaOsO$_6$ should produce some strong magnetic peaks below the (002) Bragg position. This regime corresponds to $Q$ values of ~$\sim$~1.5~\AA$^{-1}$~ or less and is characterized by a magnetic form factor for Os$^{5+}$ that is $>$~80\%\cite{11_kobayashi} of the $Q$~$=$~0 value. This discussion clearly shows that these three effects are not responsible for the very small ordered moment size and/or absence of long-range magnetic ordering in La$_2$NaOsO$_6$, and therefore a completely different factor causes this behaviour.

La$_2$NaRuO$_6$ and La$_2$NaOsO$_6$ display very unusual magnetic properties for insulating, B-site ordered double perovskite systems with magnetic atoms only occupying the B$'$ sites. The incommensurate magnetic structure of the Ru system and the drastically-reduced moment of the Os analogue are both typical features of geometrically-frustrated systems. This is puzzling when one considers that these materials are highly-distorted as the large $\beta$ values indicate, and this should relieve some of the frustration inherent to a perfect FCC lattice that the B$'$ cations decorate in the cubic double perovskites. Furthermore, most $S$~$=$~3/2 systems with dominant NN AF exchange interactions that are governed by smaller structural distortions and a higher degree of geometric frustration show Type I AF order, and so La$_2$NaRuO$_6$ and La$_2$NaOsO$_6$ may be expected to exhibit similar behaviour. However, the structural distortions also weaken the NN superexchange interactions through the B$'$-O-O-B$'$ paths, possibly to a point in La$_2$NaRuO$_6$ and La$_2$NaOsO$_6$ where the energy scale of the NN exchange approaches that of the weak NNN superexchange and then frustration arises through competition of these two different types of interactions. This alternative mechanism for frustration in the double perovskites can explain the exotic magnetic properties of La$_2$NaRuO$_6$ and La$_2$NaOsO$_6$ and has rarely played such a strong role in determining the magnetic ground state in this family of materials. The vastly different magnetic behaviour of these two systems suggests that the magnetism in this highly-distorted regime depends very sensitively on the magnitude of the structural distortion. 

In conclusion, we have investigated the magnetic properties of the highly-distorted double perovskites La$_2$NaRuO$_6$ and La$_2$NaOsO$_6$ via heat capacity and neutron powder diffraction. In contast to the Type I and II AF order most commonly found for $S$~$=$~3/2 and 5/2 double perovskite systems, our neutron diffraction results reveal an incommensurate magnetic ground state for La$_2$NaRuO$_6$ with a propagation vector of (0~0~0.091). Moreover, despite a signature of magnetic ordering from magnetic susceptibility measurements, only a broad, weak feature is observed in the specific heat and no magnetic Bragg peaks are detected for La$_2$NaOsO$_6$. This puts an upper bound on the ordered moment of 0.2~$\mu_B$. These effects are best explained by a severe weakening of the NN superexchange interactions to the point where they have a similar magnitude to the NNN exchange. Although geometric frustration is strong in ideal B-site ordered double perovskites with NN AF interactions where the magnetic atoms form an FCC lattice, this should be relieved significantly when large structural distortions are present. La$_2$NaRuO$_6$ and La$_2$NaOsO$_6$ therefore provide an alternative means to achieve frustration in double perovskite systems, namely through competing NN and NNN interactions that are delicately balanced as a result of these same structural distortions.

\begin{acknowledgments}
We acknowledge V.O. Garlea and J.E. Greedan for useful discussions. This research was supported by the US Department of Energy, Office of Basic Energy Sciences. A.A.A., C.d.l.C. and  S.E.N. were supported by the Scientific User Facilities Division, and J.-Q.Y. was supported by the Materials Science and Engineering Division. The neutron experiments were performed at the High Flux Isotope Reactor, which is sponsored by the Scientific User Facilities Division. D.E.B. and H.z.L. would like to acknowledge financial support through the Heterogeneous Functional Materials for Energy Systems (HeteroFoaM) Energy Frontiers Research Center (EFRC), funded by the US Department of Energy, Office of Basic Energy Sciences under award number DE-SC0001061.
\end{acknowledgments}


\begin{thebibliography}{99}
\bibitem{10_aharen}T. Aharen, J.E. Greedan, C.A. Bridges, A.A. Aczel, J. Rodriguez, G.J. MacDougall, G.M. Luke, T. Imai, V.K. Michaelis, S. Kroeker, H.D. Zhou, C.R. Wiebe and L.M.D. Cranswick, Phys. Rev. B {\bf 81}, 224409 (2010).
\bibitem{11_carlo}J.P. Carlo, J.P. Clancy, T. Aharen, Z. Yamani, J.P.C. Ruff, J.J. Wagman, G.J. Van Gastel, H.M.L. Noad, G.E. Granroth, J.E. Greedan, H.A. Dabkowska and B.D. Gaulin, Phys. Rev. B {\bf 84}, 100404(R), (2011).
\bibitem{10_aharen_2}T. Aharen, J.E. Greedan, C.A. Bridges, A.A. Aczel, J. Rodriguez, G.J. MacDougall, G.M. Luke, V.K. Michaelis, S. Kroeker, C.R. Wiebe, H.D. Zhou and L.M.D. Cranswick, Phys. Rev. B {\bf 81}, 064436 (2010).
\bibitem{03_wiebe}C.R. Wiebe, J.E. Greedan, P.P. Kyriakou, G.M. Luke, J.S. Gardner, A. Fukaya, I.M. Gat-Malureanu, P.L. Russo, A.T. Savici and Y.J. Uemura, Phys. Rev. B {\bf 68}, 134410 (2003).
\bibitem{02_wiebe}C.R. Wiebe, J.E. Greedan, G.M. Luke and J.S. Gardner, Phys. Rev. B {\bf 65}, 144413 (2002).
\bibitem{02_stitzer}K.E. Stitzer, M.D. Smith and H.-C. zur Loye, Solid State Science {\bf 4}, 311 (2002).
\bibitem{07_erickson}A.S. Erickson, S. Misra, G.J. Miller, R.R. Gupta, Z. Schlesinger, W.A. Harrison, J.M. Kim and I.R. Fisher, Phys. Rev. Lett. {\bf 99}, 016404 (2007).
\bibitem{10_chen}G. Chen, R. Pereira and L. Balents, Phys. Rev. B {\bf 82}, 174440 (2010).
\bibitem{10_chen_2}G. Chen and L. Balents, Phys. Rev. B {\bf 84}, 094420 (2011).
\bibitem{89_battle}P.D. Battle and C.W. Jones, Journal of Solid State Chemistry {\bf 78}, 108 (1989).
\bibitem{83_battle}P.D. Battle, J.B. Goodenough and R. Price, Journal of Solid State Chemistry {\bf 46}, 234 (1983).
\bibitem{84_battle}P.D. Battle and W.J. Macklin, Journal of Solid State Chemistry {\bf 52}, 138 (1984).
\bibitem{03_battle}P.D. Battle, C.P. Grey, M. Hervieu, C. Martin, C.A. Moore and Y. Paik, Journal of Solid State Chemistry {\bf 175}, 20 (2003).
\bibitem{06_martin}L. Ortega-San Martin, J.P. Chapman, L. Lezama, J.S. Marcos, J. Rodriguez-Fernandez, M.I. Arriortua and T. Rojo, Eur. J. Inorg. Chem. 1362 (2006). 
\bibitem{02_munoz}A. Munoz, J.A. Alonso, M.T. Casais, M.J. Martinez-Lope and M.T. Fernandez-Diaz, J. Phys. Cond. Matt. {\bf 14}, 8817 (2002).
\bibitem{04_bos}J.W.G. Bos and J.P. Attfield, Phys. Rev. B {\bf 70}, 174434 (2004).
\bibitem{10_aharen_3}T. Aharen, J.E. Greedan, F.L. Ning, T. Imai, V.K. Michaelis, S. Kroeker, H.D. Zhou, C.R. Wiebe and L.M.D. Cranswick, Phys. Rev. B {\bf 80}, 134423 (2009).
\bibitem{04_gemmill}W.R. Gemmill, M.D. Smith and H.-C. zur Loye, Journal of Solid State Chemistry {\bf 177}, 3560 (2004).
\bibitem{05_gemmill}W.R. Gemmill, M.D. Smith, R. Prozorov and H.-C. zur Loye, Inorg. Chem. {\bf  44}, 2639 (2005).
\bibitem{08_ahmed}A.S. Ahmed, H. Chen and H.K. Yuan, Phys. Stat. Sol. {\bf 245}, 720 (2008).
\bibitem{93_rodriguez}J. Rodriguez-Carvajal, Physica B {\bf 192}, 55 (1993).
\bibitem{93_andersen}M. T. Anderson, K. B. Greenwood, G. A. Taylor and K. R. Poppelmeier, Prog. Solid State Chem. {\bf 22}, 197 (1993).
\bibitem{73_khattak}C.P. Khattak, C.E. Cox and F.F.Y. Wang, AIP Conf. Proc. {\bf 10} 674 (1973).
\bibitem{11_kobayashi}K. Kobayashi, T. Nagao and M. Ito, Acta Cryst. {\bf A67}, 473 (2011).
\bibitem{09_kim}B.J. Kim, H. Jin, S.J. Moon, J.Y. Kim, B.G. Park, C.S. Leem, J. Yu, T.W. Noh, C. Kim, S.J. Oh, J.H. Park, V. Durairaj, G. Cao and E. Rotenberg, Phys. Rev. Lett. {\bf 101}, 076402 (2008).
\bibitem{12_calder}S. Calder, V.O. Garlea, D.F. McMorrow, M.D. Lumsden, M.B. Stone, J.C. Lang, J.-W. Kim, J.A. Schlueter, Y.G. Shi, K. Yamaura, Y.S. Sun, Y. Tsujimoto and A.D. Christianson, Phys. Rev. Lett. {\bf 108}, 257209 (2012).
\bibitem{12_calder_2}S. Calder, M.D. Lumsden, V.O. Garlea, J.-W. Kim, Y.G. Shi, H.L. Feng, K. Yamaura and A.D. Christianson, Phys. Rev. B {\bf 86}, 054403 (2012).

\end{thebibliography}
\end{document}